\documentclass[%
 reprint,
 amsmath,amssymb,
 aps,
]{revtex4-1}

\usepackage{dcolumn}
\usepackage{bm}

\usepackage{graphicx,capt-of}
\usepackage{color}
\usepackage{bm}
\usepackage[font=Large]{subfig}
\usepackage{tabularx,ragged2e,booktabs,caption}
\usepackage{grffile}
\usepackage{mathrsfs}
\usepackage{multirow}
\usepackage{amsmath}
\usepackage{mathtools}

\def\beq{\begin{equation}}
\def\eeq{\end{equation}}
\def\ba{\begin{eqnarray}}
\def\ea{\end{eqnarray}}

\begin{document}


\title{Charge ordering in the three-dimensional Coulomb glass at finite temperatures and low disorders}

\author{Preeti Bhandari}
\affiliation{Department of Physical Sciences, Indian Institute of Science Education and Research (IISER) Mohali, Sector 81, S.A.S. Nagar, Manauli P. O. 140306, India}
\author{Vikas Malik}%
 \email{vikasm76@gmail.com}
\affiliation{Department of Physics and Material Science, Jaypee Institute of Information Technology, Uttar Pradesh 201309, India.}

\date{\today}

\begin{abstract}
In this paper, we have studied the three dimensional Coulomb glass lattice model at half-filling using Monte Carlo Simulations. Annealing of the system shows a second-order transition from paramagnetic to charge-ordered phase for zero as well as small disorders. We have also calculated the critical exponents and transition temperature using a finite sizing scaling approach. The Monte Carlo simulation is done using the Metropolis algorithm, which allowed us to study larger lattice sizes. The transition temperature and the critical exponents at zero disorder matched the previous studies within numerical error. We found that the transition temperature of the system decreased as the disorder is increased. The values of critical exponents $\alpha$ and $\gamma$ were less and value of $\nu$ more than the corresponding zero disorder values. The use of large system sizes led to the correct variation of critical exponents with the disorder.
\end{abstract}

\keywords{Phase transitions \and Monte Carlo methods}
\maketitle


\section{\label{sec:Into}Introduction}

Many different systems, like compensated semiconductors, granular metals, ultra-thin films, etc. are modeled by the Coulomb glass (CG). The term Coulomb glass refers to systems having localized electrons (Anderson Insulators) interacting via unscreened Coulomb interaction. Transport in the CG system is via the hopping conduction mechanism. The conductivity obeys $T^{1/4}$ law given by Mott \cite{m68,m69} for temperatures where the density of states in uniform around the Fermi level. At low temperatures, there is a soft gap in the density of states which leads to $T^{1/2}$ law \cite{ab75} for conductivity.
 
The disorder manifests itself in the CG system in two ways. In the random site model, the electrons occupy random positions in the system. In the lattice model, the electrons occupy lattice sites, but the on-site energies are randomly distributed. The state of the system is thus characterized by three parameters temperature ($T$), the strength of disorder (W), and strength of interactions. For the lattice model, the on-site energies are drawn from Gaussian or box distribution. It was shown by M\"{o}bius et al. \cite{au09} that the CG system at zero disorder, shows a second-order transition from paramagnetic to antiferromagnetic phase at finite temperature in two and three dimensions. The critical exponents found were very nearly equal to the short range Ising model. 

Simulations on three dimensional ($3d$) CG, modeled using a Gaussian distribution \cite{mm09,ajmh19} have shown a phase diagram similar to the random field Ising model (RFIM). At zero temperature, there exists a critical disorder ($W_{c}$) below which one has a charge-ordered (CO) phase, and above it, one finds a disordered phase. For disorder less than the critical disorder, three dimensional CG model undergoes a second-order phase transition from the paramagnetic to CO phase at the critical temperature $T_{c}$. The finite-temperature critical exponents vary with the disorder. So like RFIM system there exist two fixed points ($W = 0$, $T = T_{c}$) and ($W = W_{c}$, $T = 0$). Both the work on CG model \cite{mm09,ajmh19} shows that the critical exponents for $W < W_{c}$ are similar to the ones at zero temperature fixed point as seen also in RFIM \cite{ad02,nv13,npii13,bjah13}. So it was concluded that at least at small disorders, both CG and RFIM belong to the same universality class   One should note that in both these works, the on-site energies are modeled by a Gaussian distribution. The numerical picture for the Gaussian $3d$ model was, to a certain extent, justified using the replica theory formalism \cite{vd07}. In the two-dimensional lattice CG model, authors \cite{pvs17} have shown that at zero temperature, there exists a first-order transition from charge-ordered phase to disordered phase as the disorder is increased. At small disorders ($W < W_{c}$), they found the second-order transition from paramagnetic to charge-ordered phase as the temperature was decreased \cite{pv19}. Coarsening dynamics of two dimensional CG model \cite{pvs19} showed remarkable similarity to the dynamics of the RFIM model \cite{lmp10,clm12}. In all these works, on-site energies were drawn from a box distribution. For disorders greater than critical disorder, especially at high disorders, many experimental and theoretical studies have claimed that the CG model exhibits glassy behavior \cite{mzm93,gdcanya97,gcdnaa98,zm97,azm98,azm00,azm02,vz04}. Mean-field studies \cite{ms07,av99,sv05,ml04} done on the three-dimensional CG lattice model at high disorders show a finite temperature glass transition using replica symmetry breaking. Numerical studies on the random site CG model have shown effects like aging, which were explained by the multi valley picture \cite{azm000,add05,mj09}. The existence of many minimas (also called pseudo ground states) for the same disorder configuration are due to the competition between disorder and long-range Coulomb interactions. In contrast, to these claims, many studies found no evidence of glass transition for disorders above $W_{c}$ \cite{mm09,bhgbg09,am10}.  

In this paper we have performed Monte Carlo simulation using Metropolis algorithm on the 3d lattice model at low disorders. Recent studies have used population annealing \cite{ajmh19} and exchange method \cite{mm09} to study the same model. Both these methods are well suited to probe for equilibrium glass transition ($W > W_{c}$). At low disorders, far below the critical disorder Metropolis algorithm sampling is an adequate method as seen in our previous work on two-dimensional CG lattice model \cite{pv19}. Our method also allowed us to study the model for much larger system sizes. This, as our results show gives a more consistent variation of the critical exponents with disorder. 

The paper is organized as follows. In Sec.\ref{M}, we discuss the Hamiltonian of the system and then in Sec.\ref{NS} our numerical simulation. The results obtained from the simulation and its interpretation are presented in Sec.\ref{reslt} and finally the conclusions are provided in Sec.\ref{conclsn}.

\section{Model}
\label{M}
 The classical three-dimensional CG lattice model \cite{ba84,mma13} can be described by the Hamiltonian
\begin{equation}
\label{HamiltonianP}
\mathcal{H} = \sum_{i=1}^N \phi_i n_i + \frac{1}{2} \sum_{i \neq j} \frac{e^{2}}{\kappa |\vec{r_i} - \vec{r_j}|} \left( n_{i} - K \right)\left( n_{j} - K \right) 
\end{equation}
where the number occupancy $n_{i}$ $\in~\{0,1\}$, the random on-site energy $\phi_{i}$ are the chosen from a uniform distribution $P(\phi)$, given by
\ba
\label{field_prob}
P(h) &=& 1/W, \quad -W/2 \leq h \leq W/2, \nonumber \\
&=& 0, \quad \mbox{otherwise} ,
\ea
here $W$ is the strength of disorder. We are concentrating only on the half-filled case which implies that the filling factor $K = 1/2$ and the model can now be mapped to an Ising model with long-ranged Antiferromagnetic interactions ($J_{ij}=e^{2}/\kappa |\vec{r_i} - \vec{r_j}|$, where $\kappa$ is the dielectric constant) using the pseudospin variable $S_i=n_i-1/2$. So the final Hamiltonian simulated by us can now be written as 
\begin{equation}
\label{Hamiltonian}
\mathcal{H} = \sum_{i=1}^N \phi_{i} S_{i} + \frac{1}{2} \sum_{i \neq j} J_{ij} S_{i} S_{j}, \quad S_i = \pm 1/2 .
\end{equation}
  
 \section{Numerical technique}
 \label{NS}
In this paper, we have performed a Monte Carlo simulation using simulated annealing on a cubic lattice ($d = 3$) to find the critical exponents of a CG lattice model. The details of the simulation are as follows: The system sizes (L) studied in this paper are $L = 4,6,8,10,12$ and $14$. We have employed periodic boundary conditions in which the simulated box is replicated in all directions and used the Ewald summation technique \cite{p21,sje80} to cope with long-range interactions. This replaces the Hamiltonian in Eq.(\ref{Hamiltonian}) in the following way:
\begin{eqnarray}
\label{EwaldH}
\mathcal{H} &=& \sum_{i=1}^N \phi_{i} S_{i} + \sum_{1 \leq i \leq j \leq N} \frac{S_{i} S_{j} \, erfc(\alpha r_{ij})}{r_{ij}} - \frac{\alpha}{\sqrt{\pi}} \sum_{i}^{N} S_{i}^{2} \nonumber \\
&+& \frac{1}{2 \pi L} \sum_{n \neq 0} \Bigg[ \frac{1}{n^{2}} exp \biggl(-\frac{\pi^{2} n^{2}}{L^{2} \alpha^{2}}\biggr) \bigg|\sum_{j=1}^{N} S_{j} exp \biggl(-\frac{2 \pi i}{L} n.r_{j}\biggr)\bigg|^{2}\Bigg] \nonumber \\
&+& \frac{2 \pi}{3 N} \bigg|\sum_{j=1}^{N} S_{j} r_{j}\bigg|^{2}
\end{eqnarray}
where
\begin{equation}
	erfc(x) = \frac{2}{\sqrt{\pi}} \int^\infty_x  e^{-t^{2}} dt	
\end{equation}
is the complementary error function \cite{mi64}. Eq.(\ref{EwaldH}) also has a regularization parameter $\alpha$, which must be chosen such that it maximizes the numerical accuracy. The real space terms in Eq.(\ref{EwaldH}) are now short-range terms, so one can easily use a spherical cut-off together with the periodic boundary conditions. All the parameters are tuned to ensure a stable convergence. In our simulation, we found that, $\alpha = 5/L$, $r_{c} = L/2$ and $n_{c} = \alpha^{2}r_{c}L/\pi$ were sufficient to get the desired results.

We found that around $W = 0.40$ the CO state was no longer the ground state for some configurations for $L \geq 8$. So we kept our calculations restricted to $W=0.20$ which was well below the start of  critical region. The Monte Carlo simulation was performed using Metropolis algorithm, details of which can be found in our earlier work on two-dimensional CG model \cite{pv19}.

\begin{figure}
\vspace*{8mm}
	\includegraphics[scale=0.32]{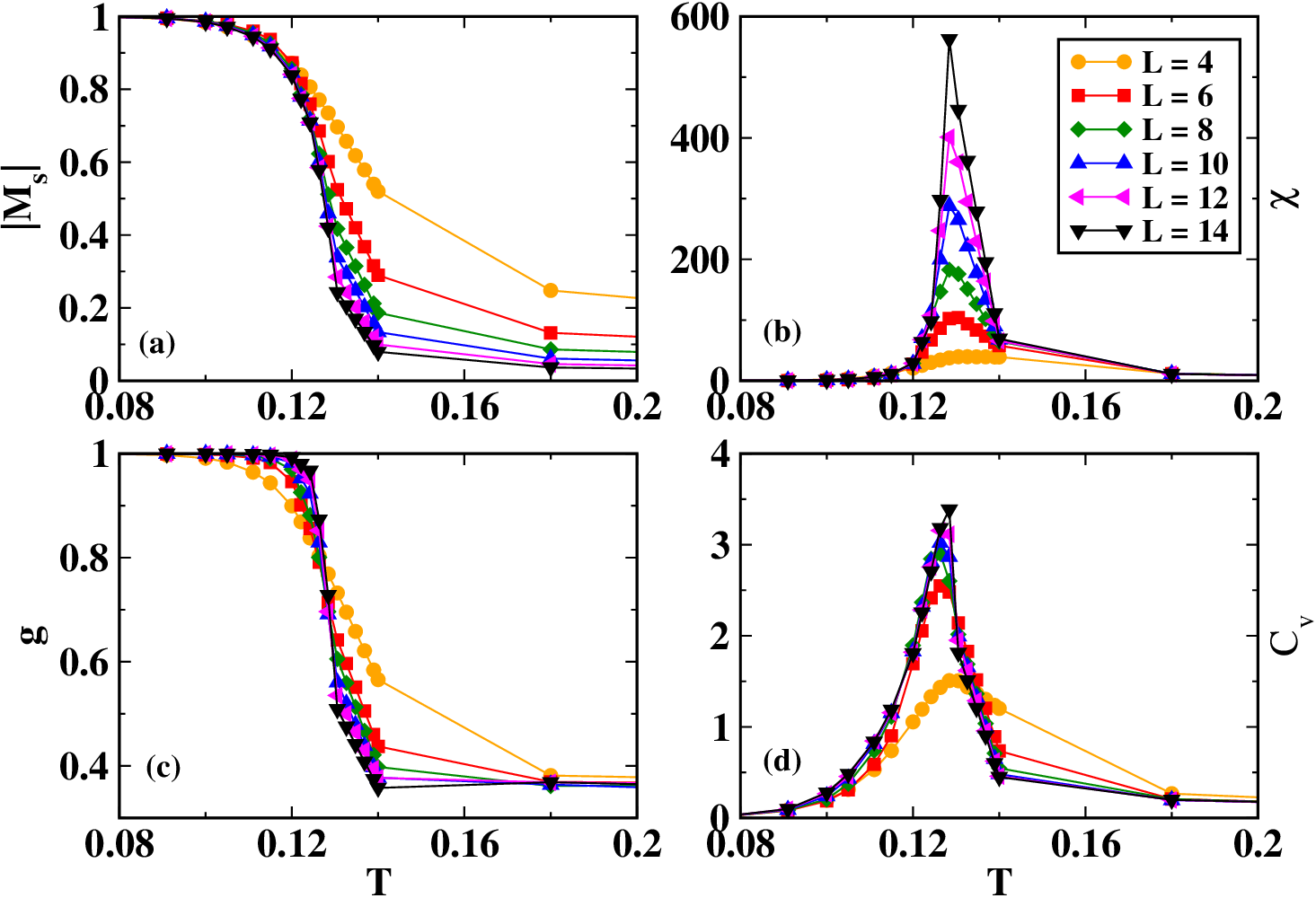}
	\caption{\label{fig:1a} (Colour online) (a) Behavior of staggered magnetization ($|M_{s}|$) as a function of temperature (T). (b) Dependence of antiferromagnetic susceptibility ($\chi$) on temperature. (c) Binder ratio ($g$) vs temperature. The crossing of this data for different system sizes gives us a value of the critical temperature, $T_{c} = 0.128$. (d) The specific heat ($C_{v}$) vs temperature. All the four graphs (a)-(d) are for disorder strength $W = 0.0$ at $L=4$ (averaged over 600 configurations), $L=6$ (averaged over 400 configurations), $L=8$ (averaged over 300 configurations), $L=10$ (averaged over 200 configurations), $L=12$ (averaged over 200 configurations) and $L=14$ (averaged over 200 configurations).}
\end{figure}

 \begin{figure}
 \vspace*{4mm}
	\includegraphics[scale=0.3]{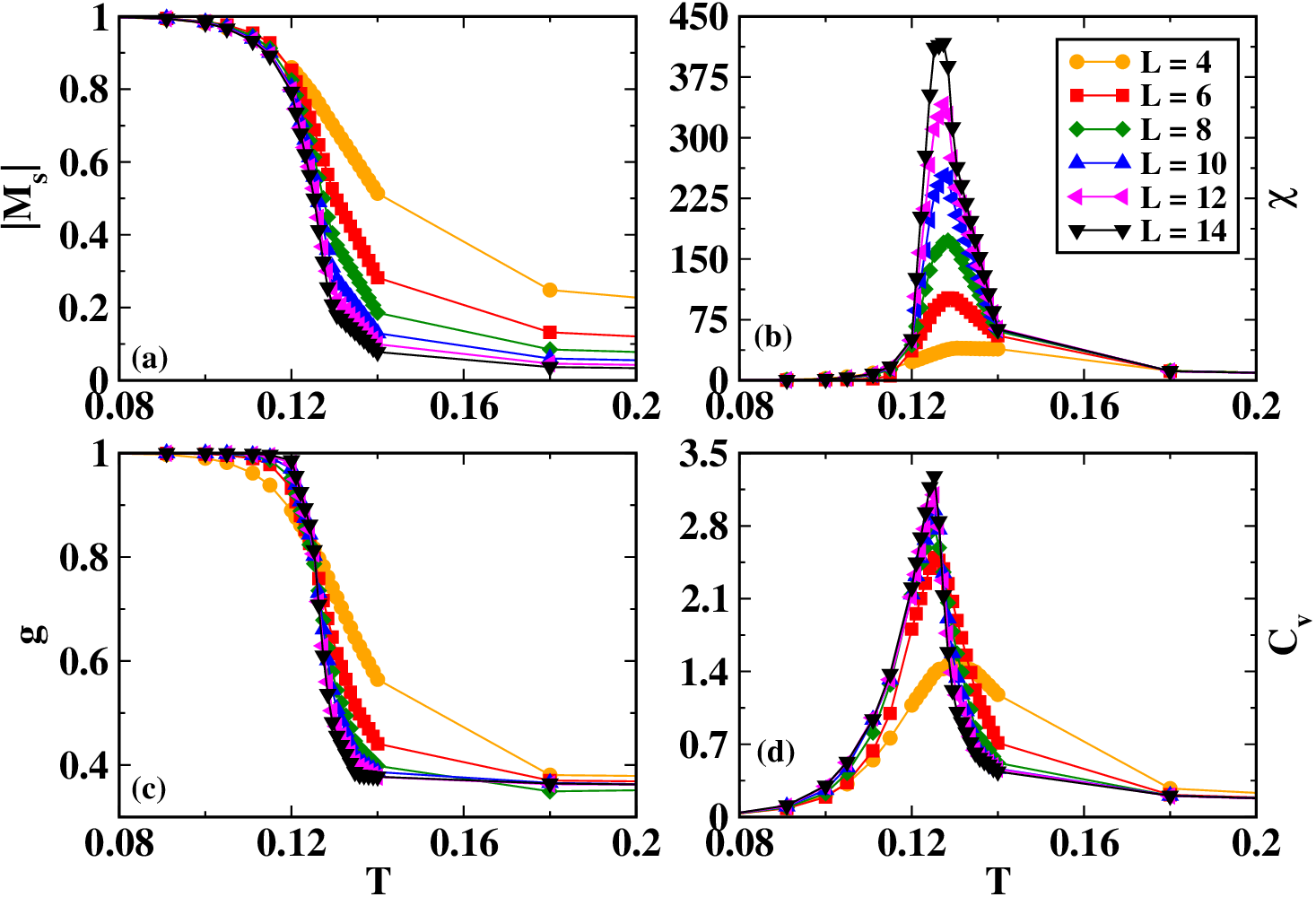}
	\caption{\label{fig:1} (Colour online) (a) Behavior of staggered magnetization ($|M_{s}|$) as a function of temperature (T). (b) Dependence of antiferromagnetic susceptibility ($\chi$) on temperature. (c) Binder ratio ($g$) vs temperature. The crossing of this data for different system sizes gives us a value of the critical temperature, $T_{c} = 0.1251$. (d) The specific heat ($C_{v}$) vs temperature. All the four graphs (a)-(d) are for disorder strength $W = 0.10$ at $L=4$ (averaged over 600 configurations), $L=6$ (averaged over 400 configurations), $L=8$ (averaged over 300 configurations), $L=10$ (averaged over 200 configurations), $L=12$ (averaged over 200 configurations) and $L=14$ (averaged over 200 configurations). }
\end{figure} 

 \begin{figure}
 \vspace*{5mm}
 	\includegraphics[scale=0.3]{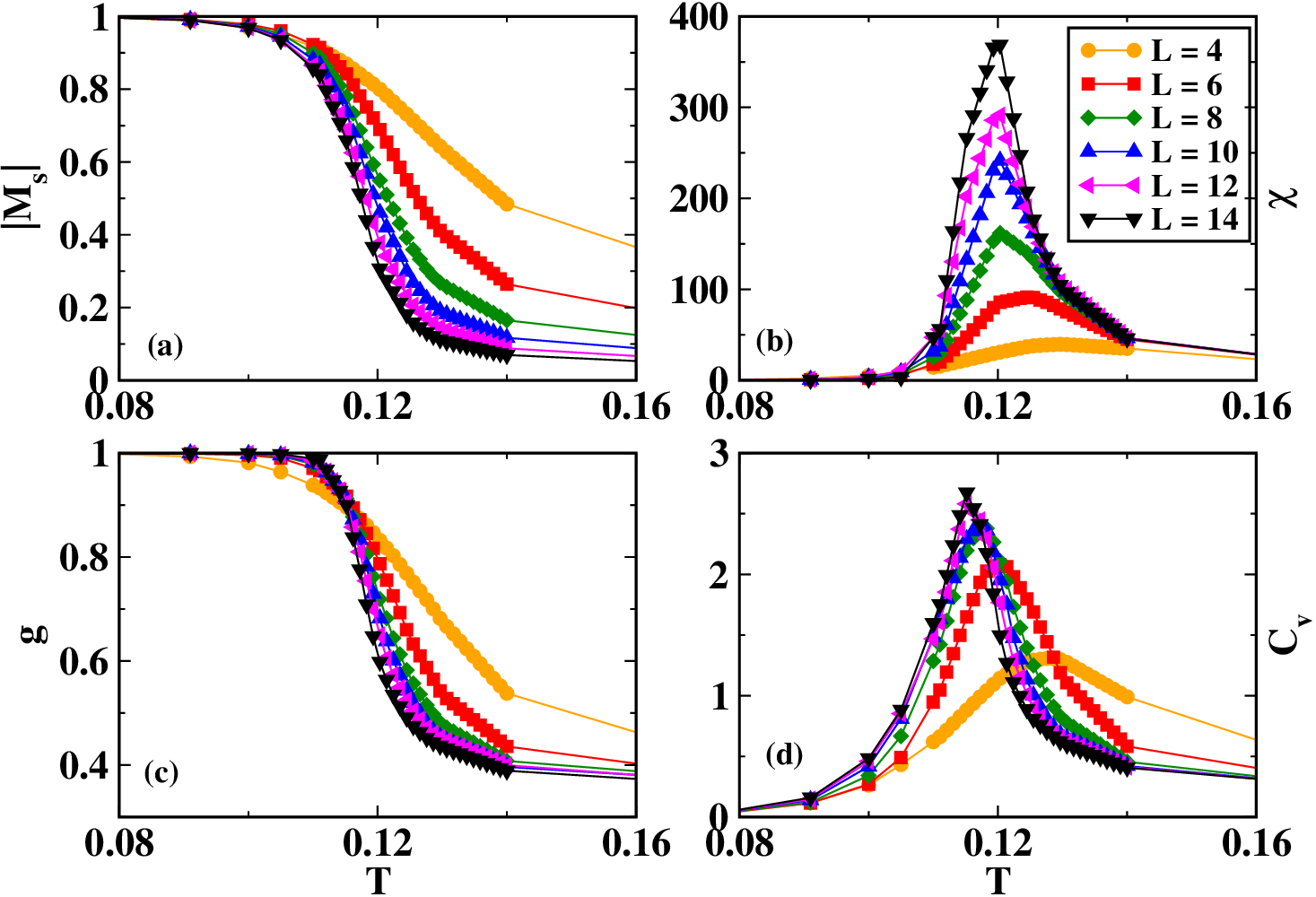}
 	\caption{\label{fig:2} (Colour online) (a) Behavior of staggered magnetization ($|M_{s}|$) as a function of temperature (T). (b) Dependence of antiferromagnetic susceptibility ($\chi$) on temperature. (c) Binder ratio ($g$) vs temperature. The crossing of this data for different system sizes gives us a value of the critical temperature, $T_{c} = 0.1143$. (d) The specific heat ($C_{v}$) vs temperature. All the four graphs (a)-(d) are for disorder strength $W = 0.20$ at $L=4$ (averaged over 600 configurations), $L=6$ (averaged over 400 configurations), $L=8$ (averaged over 300 configurations), $L=10$ (averaged over 200 configurations), $L=12$ (averaged over 200 configurations) and $L=14$ (averaged over 200 configurations). }
 \end{figure}
   
 \section{Results and discussions}
 \label{reslt}
\subsection{\textbf{Staggered magnetization}}
\label{op} The phase diagram of the three-dimensional CG model \cite{mm09,ajmh19,vd07} is quite well established now, suggesting a Charge Ordered (CO) phase below a particular value of disorder ($W_{c}$). Since the ground state is antiferromagnetic at zero and small disorders, therefore, the staggered magnetization, $M_{s}$ (which is the difference between two sublattice magnetizations) is the appropriate order parameter here. $M_{s}$, can be defined as
\begin{equation}
\label{eq8} M_{s} = \bigg[\bigg\langle\frac{1}{N}\sum^{N}_{i=1} \sigma_{i} \bigg \rangle\bigg], 
\end{equation}
where, staggered spins $\sigma_{i}$ are related to the Ising spins $S_{i}$ by the following relation:
\begin{equation}
	\sigma_{i} = (-1)^{i_{x} + i_{y} + i_{z} } S_{i}.
\end{equation}
Here, $i_{x}, i_{y}, i_{z}$ denote the $x, y, z$ coordinates of the site $i$ and $N = L^{3}$ are the total number of sites in the lattice. In Eq.(\ref{eq8}), we have used $\left\langle ... \right\rangle$ to denote the thermal average and $\left[ ... \right]$ for the ensemble average. $M_{s} \approx 0$ at high temperature where the system is in the paramagnetic phase, as the spins fluctuate randomly there. As the temperature approaches, the critical temperature spins try to align themselves in a particular order. This results into a large domain of the same $\sigma_{i}$ which finally leads to non-zero $M_{s}$. One can see this behavior of $M_{s}$ changing from $0$ to $1$ with the decrease in temperature from Fig.(\ref{fig:1a}(a)) for $W = 0.0$, Fig.(\ref{fig:1}(a)) for $W = 0.10$ and in Fig.(\ref{fig:2}(a)) for $W = 0.20$.

\subsection{\textbf{Binder ratio}}
\label{Binder} Another way to monitor phase transition is by using Binder ratio, $g$ \cite{k81}, which is defined as
\begin{equation}
\label{eqg} g = \frac{1}{2} \Bigg( 3 - \frac{[\langle M_{s}^{4}  \rangle]}{[\langle M_{s}^{2} \rangle]^{2}}\Bigg), 
\end{equation}
where $\langle M_{s}^{2} \rangle$ and $\langle M_{s}^{4}  \rangle$ denote the second and fourth moments of the staggered magnetization respectively. In the charge-ordered phase (at temperature $T < T_{c}$), $g$ approaches the value $1$, while it tends to zero at $T > T_{c}$ (paramagnetic phase). When $T = T_{c}$, $g^{*} = g$ acquires a non-trivial value, the critical Binder ratio. The behavior of $g$ vs $T$ are shown in Fig.(\ref{fig:1a}(c)), Fig.(\ref{fig:1}(c)) and in Fig.(\ref{fig:2}(c)) for $W = 0.0$, $0.10$ and $0.20$ respectively. The crossing of data at different system sizes gives the value of $T_{c} = 0.1280$ (for $W = 0.0$), $T_{c} = 0.1251$ (for $W = 0.10$) and $T_{c} = 0.1143$ (for $W = 0.20$). 

\subsection{\textbf{Susceptibility}}
\label{Xconn} Next we have calculated the antiferromagnetic susceptibility defined as follows:
\begin{equation}
\label{eq12} \chi = N \, [\langle M_{s}^{2} \rangle - \langle  M_{s} \rangle^{2}].
\end{equation}
This can be used as an additional input to analyse the phase transition in the system. In Fig.(\ref{fig:1a}(b)), Fig.(\ref{fig:1}(b)) and Fig.(\ref{fig:2}(b)) one can see that the peak in $\chi$ around $T_{c}$ becomes sharper as $L$ increases. Another important thing to note is that the distribution of $\chi$ shifts to lower towards lower temperature and also becomes broader as we increase the disorder from $W = 0.0$ to $W = 0.20$. 

\subsection{\textbf{Specific Heat}}
\label{CV} Finally, we calculate the Specific heat, $C_{v}$ as a function of temperature, using the following definition \cite{mlc04}:
\begin{equation}
\label{eq13} C_{v} = \frac{1}{N k_{B} T^{2}} \, [\langle E^{2} \rangle - \langle  E \rangle^{2}]
\end{equation}
where $E$ is the average energy per electron, and $k_{B}$ is the Boltzmann's constant. As expected the behavior of $C_{v}$ with variation in temperature and disorder (as shown in Fig.(\ref{fig:1a}(d)), Fig.(\ref{fig:1}(d)) and Fig.(\ref{fig:2}(d))) was found to be similar to what we saw in $\chi$.
 
 \begin{figure}
 \vspace*{6mm}
 	\includegraphics[scale=0.3]{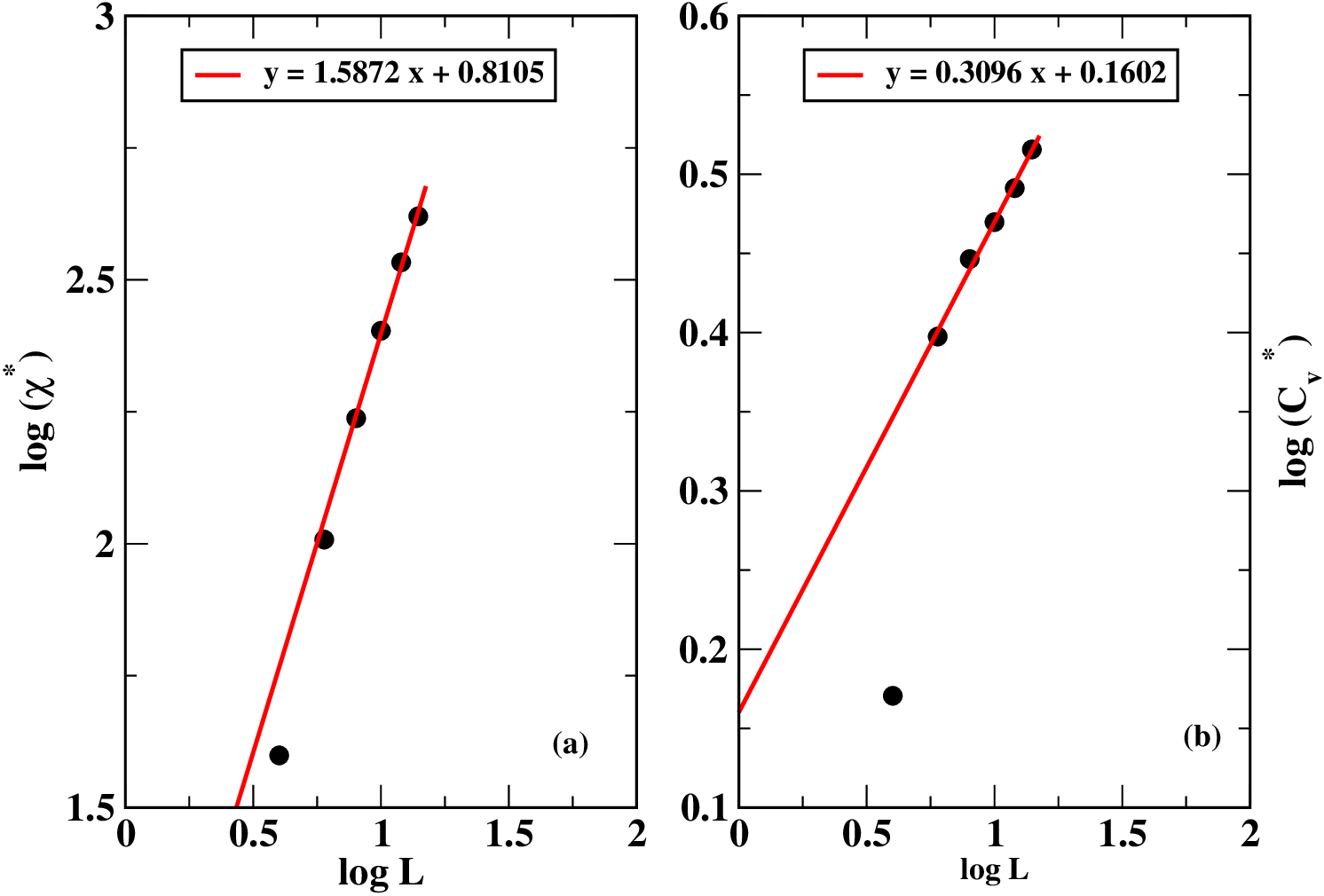}
 	\caption{\label{fig:3} (Colour online) (a) A log-log plot of the peak value of antiferromagnetic susceptibility as a function of $L$ defined as $\chi^{*}$ here vs the system size $L$. The solid line is a least-squares straight line fit giving a slope of $1.5872 \pm 0.0619$. (b) A log-log plot of the peak value of specific heat as a function of $L$ defined as $C_{v}^{*}$ here vs the system size $L$. The solid line is a least-squares straight line fit giving a slope of $0.3096 \pm 0.0162$. In both the graphs the lowest system size i.e. $L = 4$ and $6$ were not including in the fitting. The exponents were calculated for disorder strength $W = 0.10$.}
 \end{figure}  
    
 \begin{figure}
	\includegraphics[scale=0.3]{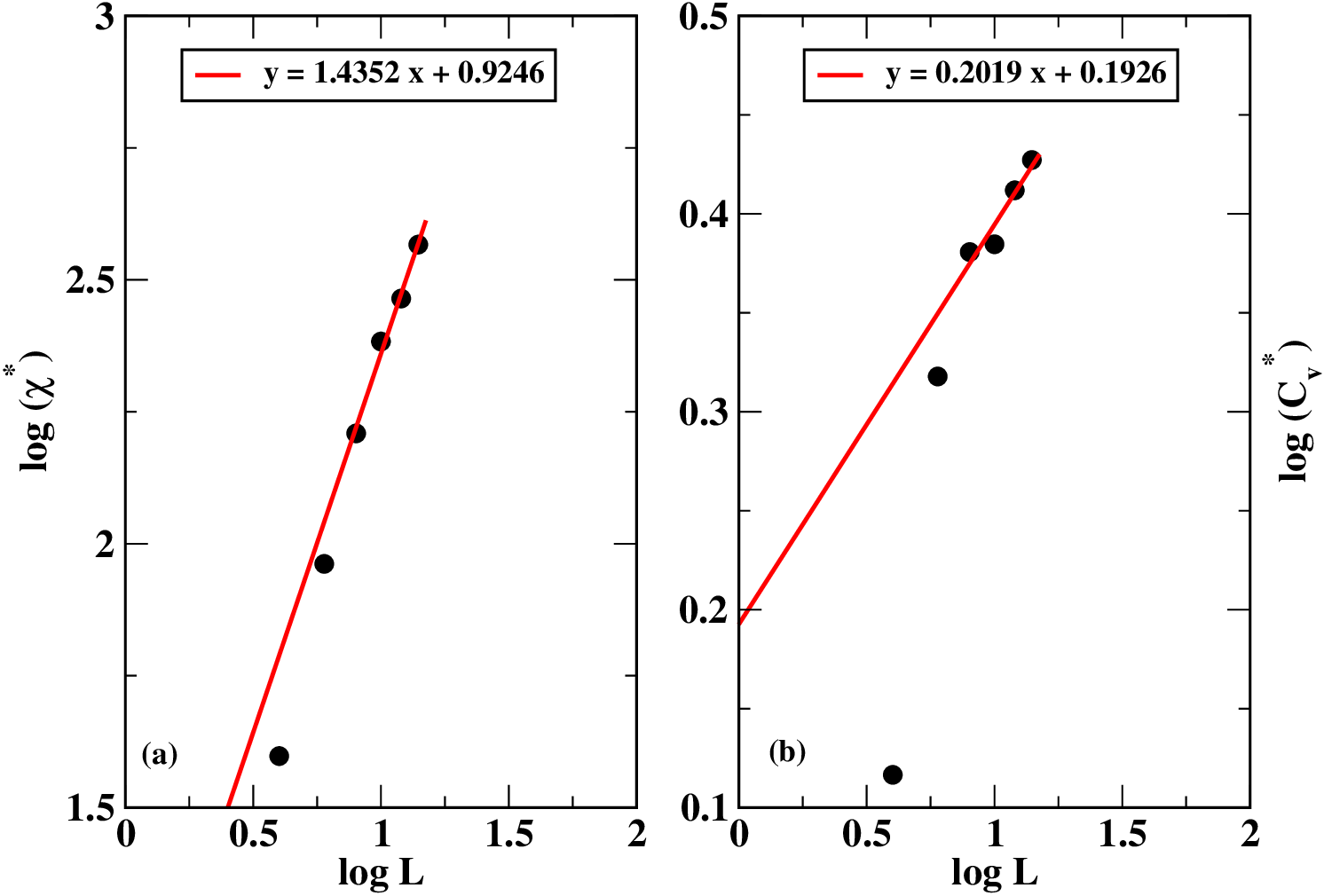}
	\vspace*{4mm}
	\caption{\label{fig:4} (Colour online) (a) A log-log plot of the peak value of antiferromagnetic susceptibility as a function of $L$ defined as $\chi^{*}$ here vs the system size $L$. The solid line is a least-squares straight line fit giving a slope of $1.4352 \pm 0.1076$. (b) A log-log plot of the peak value of specific heat as a function of $L$ defined as $C_{v}^{*}$ here vs the system size $L$. The solid line is a least-squares straight line fit giving a slope of $0.2019 \pm 0.0466$. In both the graphs the lowest system sizes i.e. $L = 4$ and $6$ were not including in the fitting. The exponents were calculated for disorder strength $W = 0.20$. }
\end{figure} 

\subsection{\textbf{Finite size scaling analysis}}
\label{fss} 
\begin{figure}
	\includegraphics[scale=0.3]{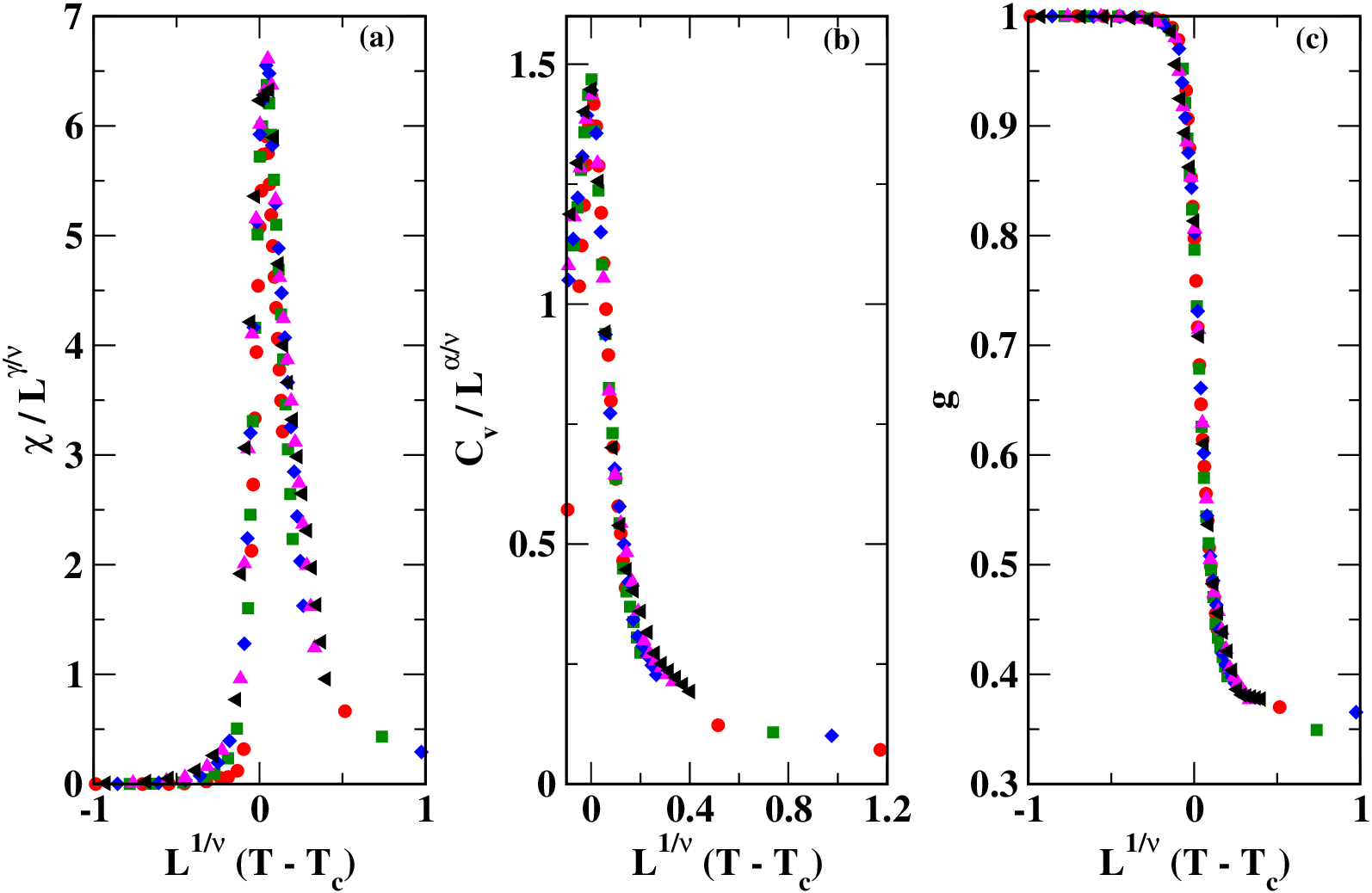}
	\caption{\label{fig:5} (Colour online) For $W = 0.10$ using the parameters $T_{c} = 0.1251$ and $\nu = 0.80$: (a) The scaling plot of the antiferromagnetic susceptibility ($\chi$) with the parameter $\gamma/\nu = 1.5872$. (b) The scaling plot of the specific heat ($C_{v}$) with the parameter $\alpha/\nu = 0.3096$. (c) The scaling plot of the Binder ratio. Data at $L=4$ is not including in any of the figures here. }
\end{figure}

\begin{figure}
	\includegraphics[scale=0.3]{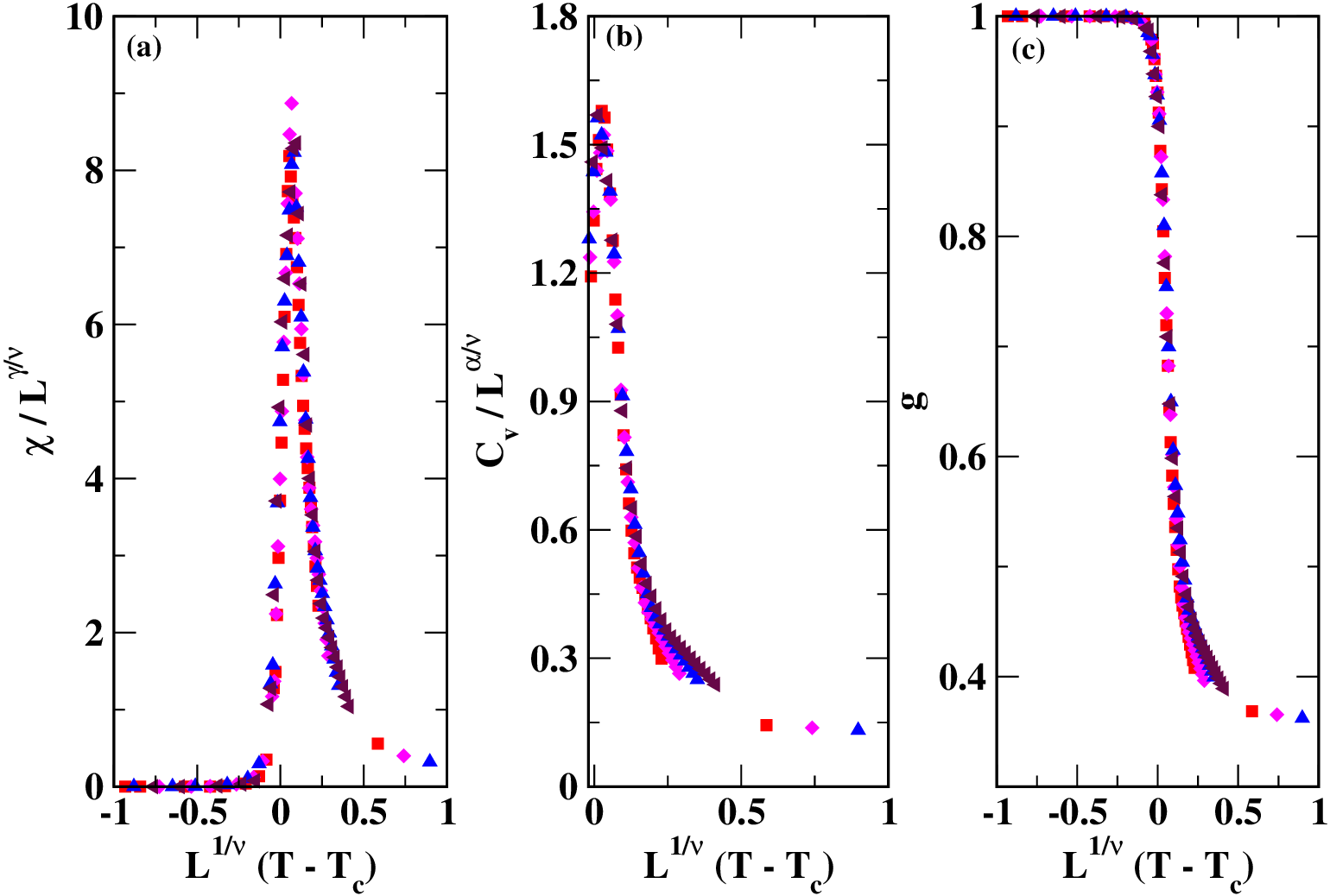}
	\caption{\label{fig:6} (Colour online) For $W = 0.20$ using the parameters $T_{c} = 0.1143$ and $\nu = 0.95$: (a) The scaling plot of the antiferromagnetic susceptibility ($\chi$) with the parameter $\gamma/\nu = 1.4352$. (b) The scaling plot of the specific heat ($C_{v}$) with the parameter $\alpha/\nu = 0.2019$. (c) The scaling plot of the Binder ratio. Data at $L=4$ is not including in any of the figures here. }
\end{figure}

For any quantity $A(T,L)$, which behaves like $A_{\infty} \sim t^{-x}$ in the infinite system when approaching the critical temperature, the finite-size scaling function is
\begin{equation}
	A(T,L) = L^{x/\nu} \, \tilde{A}(t \, L^{1/\nu})
\end{equation}
where $t = T - T_{c}$. Using the above standard finite-size scaling relation \cite{h95}, we get
\begin{equation}
\label{geq} g (T,L) \approx \hspace*{1mm} \tilde{g} \hspace*{1mm} [L^{1/\nu} (T-T_{c})]
\end{equation}
\begin{equation}
\label{Xeq} \chi (T,L) \approx L^{\gamma/\nu} \hspace*{1mm} \tilde{\chi} \hspace*{1mm} [L^{1/\nu} (T-T_{c})]
\end{equation}
\begin{equation}
\label{Ceq} C_{v} (T,L) \approx L^{\alpha/\nu} \hspace*{1mm} \tilde{C_{v}} \hspace*{1mm} [L^{1/\nu} (T-T_{c})]
\end{equation}
Now, the critical exponents $\alpha/\nu$ and $\gamma/\nu$ can be extracted by plotting a least-squares straight-line fit of $C_{v}^{*}$ and $\chi^{*}$ versus $L$ respectively in a log-log plot (as shown in Fig.(\ref{fig:3}, \ref{fig:4}). Here, $C_{v}^{*}$ and $\chi^{*}$ are the peak values of $C_{v}$ and $\chi$ at the each system size. The critical exponent of correlation length $\nu$ was used as an adjustable parameter to collapse the data of $g$ (see Fig.(\ref{fig:5}(c), \ref{fig:6}(c)). The value of $T_{c}$, $\nu$, $\gamma/\nu$ and $\alpha/\nu$ at different disorders are summarized in Table.\ref{T1}. For comparison we have also included the critical exponents at $W = 0$, which are close to one calculated by other authors \cite{ajmh19}, claiming $T_{c} = 0.1280$, $\gamma/\nu = 2.05$, $\nu = 0.76$ and $\alpha/\nu = 0.550$. The final scaled plots of $\chi$, $C_{v}$ and $g$ are shown in Fig.(\ref{fig:5}, \ref{fig:6}).

\begin{table}
\caption{Critical exponents for the three-dimensional lattice Coulomb glass model.}
\label{T1}       
\center
\begin{tabular}{lllll}
\hline\noalign{\smallskip}
$W$ & $T_{c}$ & $\gamma/\nu$ & $\nu$ & $\alpha/\nu$ \\
\noalign{\smallskip}\hline\noalign{\smallskip}
0.00 & 0.1280 & 1.9633 $\pm$ 0.0853 & 0.70 & 0.5939 $\pm$ 0.1020 \\ 
0.10 & 0.1251 & 1.5872 $\pm$ 0.0619 & 0.80 & 0.3096 $\pm$ 0.0162\\ 
0.20 & 0.1143 & 1.4352 $\pm$ 0.1076 & 0.95 & 0.2019 $\pm$ 0.0466\\
\noalign{\smallskip}\hline
\end{tabular}
\end{table}

 \section{Conclusions}
 \label{conclsn}
 We have studied the critical behavior of a three-dimensional Coulomb Glass lattice model at half-filling using Monte Carlo simulation via simulated annealing at zero and small disorders. The summary of our results is as follows.

(a) We found a finite-temperature phase transition from a high-temperature paramagnetic phase to a low-temperature charge-ordered phase for zero disorder and disorders less than $W_{c}$.

(b) Finite-size scaling analysis was done to determine the transition temperature ($T_{c}$), and the critical exponents of correlation length ($\nu$), susceptibility ($\gamma/\nu$) and specific heat ($\alpha/\nu$). We found that $T_{c}, \gamma/\nu$ and $\alpha/\nu$ decreases as the disorder ($W$) increases but $\nu$ increases with increase in $W$. The correct variations in the critical exponents with disorder have been achieved because of the use of larger system sizes. Our estimates of critical exponents at zero disorder are close to the one obtained by Amin et al. \cite{ajmh19}.

(c) We also found that the peak value of susceptibility and specific heat decreases, and the distribution gets broader with an increase in disorder.
 
 \section*{ACKNOWLEDGEMENT}
 We thank late Professor Deepak Kumar for useful discussions on the subject. We wish to thank NMEICT cloud service provided by BAADAL team, cloud computing platform, IIT Delhi for the computational facility. PB gratefully acknowledges IISER Mohali for the financial support and for the use of High Performance Computing
Facility.


\begin{thebibliography}{99}

 \bibitem{m68}
  N. F. Mott, J. J. Non-Cryst. Solids \textbf{1}, 1 (1968).

 \bibitem{m69}
  N. F. Mott, Phil. Mag. B \textbf{19}, 835 (1969).
  
 \bibitem{ab75}
  A. L. Efros and B. I. Shklovskii, J. Phys. C: Solid State Phys. \textbf{8}, L49 (1975).   
  
 \bibitem{au09}      
  A. M\"{o}bius, and U. K. R\"{o}ssler, Phys. Rev. B \textbf{79}, 174206 (2009).   
  
 \bibitem{mm09}      
  M. Goethe, and M. Palassini, Phys. Rev. Lett. \textbf{103}, 045702 (2009).  

 \bibitem{ajmh19}
  A. Barzegar, J. C. Andresen, M. Schechter, and H. G. Katzgraber,  Phys. Rev. B {\bf 100}, 104418 (2019). 
 
 \bibitem{ad02}  
  A. A. Middleton and D. S. Fisher, Phys. Rev. B \textbf{65}, 134411 (2002).   
  
 \bibitem{nv13}  
  N. G. Fytas and V. Martin-Mayor, Phys. Rev. Lett. \textbf{110}, 227201 (2013).
  
 \bibitem{npii13} 
  N. G. Fytas, P. E. Theodorakis, I. Georgiou, and I. Lelidis, Eur. Phys. J.
  B \textbf{86}, 268 (2013). 
  
 \bibitem{bjah13}
  B. Ahrens, J. Xiao, A. K. Hartmann, and H. G. Katzgraber, Phys. Rev. B \textbf{88}, 174408 (2013).  
  
 \bibitem{vd07}
  V. Malik and D. Kumar, Phys. Rev. B \textbf{76}, 125207 (2007).    

 \bibitem{pvs17}                 
  P. Bhandari, V. Malik, and S. R. Ahmad, Phys. Rev. B \textbf{95}, 184203 (2017).
  
 \bibitem{pv19}
  P. Bhandari and V. Malik, Eur. Phys. J. B \textbf{92}, 147 (2019).  
  
 \bibitem{pvs19}                 
  P. Bhandari, V. Malik, and S. Puri, Phys. Rev. E \textbf{99}, 052113 (2019).    

 \bibitem{lmp10}
  E. Lippiello, A. Mukherjee, S. Puri and M. Zannetti, Europhys. Lett. \textbf{90}, 46006 (2010);
  F. Corberi, E. Lippiello, A. Mukherjee, S. Puri and M. Zannetti, J. Stat. Mech. (2011) P03016.

 \bibitem{clm12}
  F. Corberi, E. Lippiello, A. Mukherjee, S. Puri and M. Zannetti, Phys. Rev. E, \textbf{85} 021141 (2012).  
  
 \bibitem{ms07}
  M. M\"{u}ller and S. Pankov, Phys. Rev. B \textbf{75}, 144201 (2007).   
  
 \bibitem{av99}
  A. A. Pastor and V. Dobrosavljevi\'{c}, Phys. Rev. Lett. \textbf{83}, 4642 (1999).

 \bibitem{sv05}
  S. Pankov and V. Dobrosavljevi\'{c}, Phys. Rev. Lett. \textbf{94}, 046402 (2005).  

 \bibitem{ml04}
  M. M\"{u}ller and L. B. Ioffe, Phys. Rev. Lett. \textbf{93}, 256403 (2004).
  
 \bibitem{mzm93}
  M. Ben-Chorin, Z. Ovadyahu, and M. Pollak, Phys. Rev. B \textbf{48}, 15025 (1993).

 \bibitem{gdcanya97}
  G. Martinez-Arizala, D. E. Grupp, C. Christiansen, A. Mack, N. Markovic, Y. Seguchi, and A. M. Goldman, Phys. Rev. Lett. \textbf{78}, 1130 (1997)  

 \bibitem{gcdnaa98}
  G. Martinez-Arizala, C. Christiansen, D. E. Grupp, N. Markovic, A. Mack, and A. M. Goldman, Phys. Rev. B \textbf{57}, R670 (1998). 

 \bibitem{zm97}
  Z. Ovadyahu and M. Pollak, Phys. Rev. Lett. \textbf{79}, 459 (1997) 

 \bibitem{azm98}
  A. Vaknin, Z. Ovadyahu, and M. Pollak, Phys. Rev. Lett. \textbf{81}, 669 (1998).

 \bibitem{azm00}
  A. Vaknin, Z. Ovadyahu and M. Pollak, Phys. Rev. Lett. \textbf{84}, 3402 (2000).

 \bibitem{azm02}
  A. Vaknin, Z. Ovadyahu and M. Pollak, Phys. Rev. B. \textbf{65}, 134208 (2002). 

 \bibitem{vz04}
  V. Orlyanchik and Z. Ovadyahu, Phys. Rev. Lett. \textbf{92}, 066801 (2004). 
  
 \bibitem{azm000}
  A. Vaknin, Z. Ovadyahu and M. Pollak, Phys. Rev. B. \textbf{61}, 6692 (2000).  

 \bibitem{add05}
  A. B. Kolton, D. R. Grempel and D. Dom\'{i}nguez, Phys. Rev. B. \textbf{71}, 024206 (2005).

 \bibitem{mj09}
  M. Kirkengen and J. Bergli, Phys. Rev. B. \textbf{79}, 075205 (2009).     
  
 \bibitem{bhgbg09}
  B. Surer, H. G. Katzgraber, G. T. Zimanyi, B. A. Allgood and G. Blatter, Phys. Rev. Lett. \textbf{102}, 067205 (2009).

 \bibitem{am10}
  A. M\"{o}bius and M. Richter, Phys. Rev. Lett. \textbf{105}, 039701 (2010).  
  
 \bibitem{ba84}
  B. I. Shklovskii and A. L. Efros, Electronic Properties of Doped Semiconductors, Heidelberg: Springer, Heidelberg, (1984). 
  
 \bibitem{mma13}
  M. Pollak, M. Ortu\~{n}o, and A. Frydman, The Electron Glass, Cambridge University Press, New York, (2013).  
  
 \bibitem{p21}
  P. P. Ewald, Ann. Phys. \textbf{369}, 253 (1921).

 \bibitem{sje80} 
  S. W. de Leeuw, J. W. Perram, and E. R. Smith, Proc. R. Soc. A \textbf{373}, 27 (1980).

 \bibitem{mi64}
  M. Abramowitz and I. A. Stegun, \textit{Handbook of Mathematical Functions with Formulas, Graphs, and Mathematical Tables} (Dover, New York, 1964).

 \bibitem{k81}
  K. Binder, Phys. Rev. Lett. \textbf{47}, 693 (1981). 
  
 \bibitem{mlc04}    
  M. H. Overlin, L. A. Wong, and C. C. Yu, Phys. Rev. B \textbf{70}, 214203 (2004).     
  
 \bibitem{h95}
  H. Rieger, Phys. Rev. B \textbf{52}, 6659 (1995).    
                                                        
\end{thebibliography}
\end{document}